\documentclass[twocolumn,showpacs,showkeys,preprintnumbers,prd,nofootinbib]{revtex4-1}

\usepackage{amsmath}
\usepackage{amsfonts}
\usepackage{amssymb}
\usepackage{graphicx}
\usepackage{color}
\usepackage[colorlinks=true,citecolor=blue]{hyperref}
\usepackage{booktabs}
\usepackage{cleveref}

\Crefname{equation}{Eq.}{Eqs.}
\Crefname{figure}{Fig.}{Figs.}
\Crefname{section}{Sec.}{Secs.}

\usepackage{etoolbox}
\makeatletter
\appto{\appendix}{%
  \@ifstar{\def\theequation@prefix{A.}}%
          {}%
}
\makeatother

\begin{document}

\title{Holographic dark energy from nonadditive entropy: cosmological perturbations and observational constraints}

\author{Rocco D'Agostino}
\email{rocco.dagostino@roma2.infn.it}
\affiliation{Istituto Nazionale di Fisica Nucleare (INFN), Sez. di Roma ``Tor Vergata'', Via della Ricerca Scientifica 1, I-00133, Roma, Italy.}

\begin{abstract}

We apply the holographic principle in the cosmological context through the nonadditive Tsallis entropy, used to describe the thermodynamic properties of nonstandard statistical systems such as the gravitational ones. 
Assuming the future event horizon as the infrared cutoff, we build a dark energy model free from cosmological inconsistencies, which includes standard thermodynamics and standard holographic dark energy as a limiting case. We thus describe the dynamics of Tsallis holographic dark energy in a flat FLRW background. 
Hence, we investigate cosmological perturbations in the linear regime on sub-horizon scales.
We study the growth of matter fluctuations in the case of clustering dark matter and a homogeneous dark energy component.
Furthermore, we employ the most recent late-time cosmic data to test the observational viability of our theoretical scenario.  
We thus obtain constraints on the free parameters of the model by means of Monte Carlo numerical method. 
We also used Bayesian selection criteria to estimate the statistical preference for Tsallis holographic dark energy compared to the concordance $\Lambda$CDM paradigm.
Our results show deviations from standard holographic dark energy within the $2\sigma$ confidence level. 
Finally, the analysis of the dark energy equation of state indicates a quintessence-like behaviour with no evidence for phantom-divide crossing at the $1\sigma$ level.

\end{abstract}

\maketitle

\section{Introduction}

The accelerated phase of the late-time cosmic expansion first observed in the Hubble diagram of type Ia Supernovae \cite{Riess98,Perlmutter99} cannot be explained by assuming matter (baryons + cold dark matter) and radiation as the only constituents in the energy budget of the universe. 
Modifications of general relativity \cite{Wands94,Capozziello02,Nojiri03,Carroll04,Cai16,Nojiri17,Capozziello18,Abedi18,DAgostino18,Capozziello19,review} represent one possibility to address the late-time acceleration problem considered to be of gravitational origin. 
On the other hand, the observed behaviour of the cosmic fluid can be attributed to the extra degrees of freedom of new exotic terms in the energy-momentum tensor, giving rise to the so-called \emph{dark energy} models \cite{Sahni00,Huterer01,Padmanabhan03,Copeland06,Aviles12}. 
The cosmological constant $\Lambda$, while the most simple dark energy candidate, does not provide a satisfactory solution of the issue  due to the fine-tuning and coincidence problems \cite{Weinberg89,Zlatev99,Carroll01,Sahni02}.
Also, dynamical dark energy models characterized by a time-evolving equation of state are purely based on phenomenological arguments, which make them unlikely represent an effective solution to the cosmological puzzle \cite{Peebles-Ratra88,Caldwell98,Armendariz00}.

Alternatively, the origin and nature of dark energy can be studied through the \emph{holographic principle} of quantum gravity \cite{Hooft93} applied in the cosmological context.  In this scheme, the vacuum energy from the ultraviolet quantum cutoff is related to the characteristic length of the universe, and all the physical degrees of freedom are described in terms of some quantities at the universe's boundary \cite{Cohen99}.
The resulting holographic dark energy (HDE) models can explain the current acceleration and are found to be in agreement with observations \cite{Zhang05,Feng07,Li09,Luongo17,Malekjani18}.
These scenarios provide also interesting cosmological features being able to successfully alleviate the fine-tuning and coincidence problems \cite{Li04,Horvat04,Pavon05,Nojiri06}. 
Moreover, the holographic principle has been invoked to unify dark matter and dark energy into a single scheme by relating the scale length to second-order curvature invariants \cite{Aviles11}.

Similarly to a black hole, in the cosmological applications of holography the entropy of the whole universe is proportional to its area. 
However, the standard Boltzmann-Gibbs (BG) theory is not valid for gravitational systems, where the partition function diverges \cite{Gibbs1902}. The fundamental hypothesis of the BG entropy is the weak probabilistic correlations between the elements of the system. The BG entropy is assumed to be an additive function for two statistically independent systems.
However, systems with long-range interactions, such as gravitational ones, are \emph{nonadditive} since the energy between the different parts of the system is not negligible compared to the total energy.
In this case, the unusual thermodynamic properties require the use of a generalized formalism known as Tsallis entropy \cite{Tsallis88,Lyra98,Tsallis13}, parametrized by a nonadditive exponent $\beta$.
This generalized approach reduces to the standard BG entropy in the limit $\beta=1$. 

Dark energy in the framework of Tsallis statistics was first investigated in \cite{Nunes}. 
Later on,  a Tsallis holographic dark energy (THDE) model was proposed in \cite{Tavayef18}, where standard entropy and usual HDE are not accounted as limiting case. 
This disadvantage arises from considering the Hubble horizon as the characteristic length of the universe, which leads to unrealistic cosmological scenarios in the case of standard HDE \cite{Hsu04,Jahromi18}.
More recently, the Tsallis entropy has been considered to investigate the variational behaviour of the nonadditive exponent with the energy scale \cite{Nojiri19}.

In this work, we assume the future event horizon as the universe's characteristic length.
This permits a consistent generalization of additive entropy which recovers standard HDE in the limit $\beta=1$.

The structure of the paper is as follows.
In \Cref{sec:Tsallis}, we present the dark energy model built upon the cosmological application of the holographic principle through the nonadditive Tsallis entropy.  
In \Cref{sec:background}, we derive the dynamical equations governing the background evolution of a flat Friedmann-Lema\^itre-Roberson-Walker (FLRW) universe described by the THDE model. 
Then, in \Cref{sec:perturbations} we discuss cosmological perturbations of the fluid composed by matter and HDE in the linear regime on sub-horizon scales.
In \Cref{sec:datasets}, we describe the experimental datasets we employ to test our theoretical scenario.
In \Cref{sec:constraints}, we implement a Bayesian Monte Carlo approach to get bounds on the cosmological parameters, and we study the statistical performance of the THDE models through model selection criteria.
Finally, in \Cref{sec:conclusion} we summarize the obtained results and discuss the future perspectives of our work. 

Throughout this paper, we use physical units such that $\hbar=k_B=c=1$.

\section{Tsallis holographic dark energy}
\label{sec:Tsallis}

The standard derivation of the HDE density is based on the entropy-area relation of black holes, $S\propto A$, being $A=4\pi L^2$ the area of the horizon \cite{Cohen99}.
Quantum gravity considerations, however, show that the above definition can be actually modified \cite{Wang16}.
In fact, a generalization of the BG theory resulting from the application of nonextensive statistical mechanics leads to the definition of the Tsallis entropy \cite{Tsallis88}:
\begin{equation}
S_T=\gamma A^\beta\ ,
\label{Tsallis entropy}
\end{equation}
where $A\propto L^2$ is the area of a $d$-dimensional system with characteristic length $L$.
Here, $\gamma$ is an unknown constant and $\beta=d/(d-1)$ for $d>1$, under the hypothesis of equal probabilities.
\Cref{Tsallis entropy} reduces to the additive Bekenstein entropy for $\beta=1$ and $\gamma=2\pi M_\text{P}^2$, where $M_\text{P}=(8\pi G)^{-1/2}$ is the reduced Planck mass. 

The holographic principle states that all the degrees of freedom of a physical system can be projected onto its boundary \cite{Hooft93}. 
Based on this argument, it was proposed that the entropy of the system is related to the infrared cutoff $L$ and the ultraviolet cutoff $\Lambda$  as \cite{Cohen99}
\begin{equation}
L^3\Lambda^3\leq S^{3/4}\ ,
\end{equation}
which can be combined with \Cref{Tsallis entropy} to obtain
\begin{equation}
\Lambda^4\leq(4\pi)^\beta \gamma L^{2\beta-4}\ .
\end{equation}
Under the holographic hypothesis, $\Lambda^4$ thus represents the THDE density which reads \cite{Guberina07,Ghaffari15}
\begin{equation}
\rho_{de}=BL^{2\beta-4}\ ,
\label{eq:rho_de}
\end{equation}
with $B=3c^2M_\text{P}^2$, where $c^2$ is a dimensionless quantity, usually assumed to be constant \cite{Radicella10}.
We note that standard HDE is included as the sub-case $\beta=1$, while \Cref{eq:rho_de} gives the standard cosmological constant for $\beta=2$.

To study the cosmological dynamics of THDE, we consider a flat FLRW metric:
\begin{equation}
ds^2=-dt^2+a^2(t)\delta_{ij}dx^idx^j\ ,
\label{FLRW}
\end{equation}
where $a(t)$ is the scale factor such that $a(t_0)=1$ at the present time. 
In the formulation of a HDE model, one needs to identify the largest length $L$ of the theory.
The model recently proposed in \cite{Tavayef18} considers the Hubble horizon $H^{-1}$ playing the role of $L$ in \Cref{eq:rho_de}, where $H\equiv \dot{a}/a$ is the Hubble parameter.
Unfortunately, as shown in  \cite{Hsu04}, this choice leads to cosmological inconsistencies in the case of standard HDE models.
For this reason, the resulting model does not include standard thermodynamics and standard HDE as sub-classes. 

A remedy for this drawback is to consider the future event horizon \cite{Saridakis18}:
\begin{equation}
R_h=a\int_t^\infty\dfrac{dt}{a}=a\int_a^\infty\dfrac{da}{Ha^2}\ ,
\label{eq:future event}
\end{equation}
which can be used to build a consistent THDE model. One thus obtains
\begin{equation}
\rho_{de}=BR_h^{2\beta-4}\ .
\label{eq:rho_de2}
\end{equation}
In what follows, we focus on the case $\beta\neq2$ to explore dynamics beyond the cosmological constant scenario.

\section{Background evolution}
\label{sec:background}

For a homogeneous and isotropic universe described by the metric (\ref{FLRW}), filled with a perfect fluid of pressureless matter and  a dark energy component, the Friedmann equation take the form
\begin{align}
&H^2=\dfrac{1}{3M^2_\text{P}}(\rho_m+\rho_{de})\ \label{eq:first Friedmann} ,  \\
&\dot{H}=-\dfrac{1}{2M^2_\text{P}}(\rho_m+\rho_{de}+p_{de})\ ,
\end{align}
where the `dot' denotes derivative with respect to the cosmic time. Here, $\rho_m$ is the matter energy density, while $\rho_{de}$ and $p_{de}$ are the dark energy density and pressure, respectively.
Introducing the critical density $\rho_c=3M_\text{P}^2H^2$, one can define the normalized density parameters of the cosmic species $\Omega_i\equiv \rho_i/\rho_c$:
\begin{align}
&\Omega_m=\dfrac{\rho_m}{3M_\text{P}^2H^2}\  \label{eq:Omega_m},  \\
&\Omega_{de}=\dfrac{\rho_{de}}{3M_\text{P}^2H^2}\ ,
\label{eq:Omega_de}
\end{align}
which satisfy $\Omega_m+\Omega_{de}=1$. Using the above definitions and combining \Cref{eq:future event,eq:rho_de2}, one obtains
\begin{equation}
\int_x^\infty\dfrac{dx}{Ha}=\dfrac{1}{a}\left(\dfrac{B}{3M_\text{P}^2H^2\Omega_{de}}\right)^{1/(4-2\beta)}\ ,
\label{eq:integral}
\end{equation}
where we have introduced the variable $x\equiv\ln a$.
Assuming no interaction between the cosmic sectors,  the matter conservation equation reads
\begin{equation}
\dot{\rho}_m+3H\rho_m=0\ ,
\end{equation}
which gives $\rho_m=\rho_{m0}a^{-3}$, where the subscript `0' denotes the value of a quantity at the present time. 
Then, \Cref{eq:Omega_m} becomes $\Omega_{m}=\Omega_{m0}H_0^2/(a^3H^2)$ and one can rewrite \Cref{eq:first Friedmann} as 
\begin{equation}
\dfrac{1}{Ha}=\dfrac{1}{H_0}\sqrt{\dfrac{a(1-\Omega_{de})}{\Omega_{m0}}}\ .
\label{eq:1/Ha}
\end{equation}
Plugging this result into \Cref{eq:integral} and differentiating with respect to $x$, we obtain the equation describing the evolution of THDE:
\begin{equation}
\dfrac{d\Omega_{de}/dx}{\Omega_{de}(1-\Omega_{de})}=2\beta-1+\eta(1-\Omega_{de})^{\frac{1-\beta}{2(2-\beta)}}\Omega_{de}^{\frac{1}{2(2-\beta)}}e^{\frac{3(1-\beta)}{2(2-\beta)}x},
\label{eq:evolution THDE}
\end{equation}
where 
\begin{equation}
\eta\equiv 2(2-\beta)\left(H_0\sqrt{\Omega_{m0}}\right)^\frac{1-\beta}{\beta-2}\left(\dfrac{B}{3M_\text{P}^2}\right)^{\frac{1}{2(2-\beta)}}\ .
\end{equation}
We note that, in the limit $\beta=1$, \Cref{eq:evolution THDE} possesses an analytical solution which coincides with the usual HDE \cite{Li04}. For $\beta\neq 1$, it cannot be solved analytically and only a numerical approach is possible.

On the other hand, the evolution of the THDE equation of state parameter $w_{de}\equiv p_{de}/\rho_{de}$ can be obtained from the following conservation equation:
\begin{equation}
\dot{\rho}_{de}+3H\rho_{de}(1+w_{de})=0\ .
\label{eq:continuity DE}
\end{equation}
Differentiating \Cref{eq:rho_de2} with the help of \Cref{eq:future event}, we get
\begin{equation}
\dot{\rho}_{de}=2(\beta-2)BR_h^{2\beta-5}(HR_h-1)\ ,
\end{equation}
and using \Cref{eq:rho_de2} to eliminate $R_h$, from \Cref{eq:continuity DE} one finds
\begin{align}
&2(\beta-2)B\left(\dfrac{\rho_{de}}{B}\right)^\frac{2\beta-5}{2(\beta-2)}\left[H\left(\dfrac{\rho_{de}}{B}\right)^\frac{1}{2(\beta-2)}-1\right] \nonumber \\
&+3H\rho_{de}(1+w_{de})=0\ .
\end{align}
Hence, making use of \Cref{eq:Omega_de,eq:1/Ha}  to substitute $\rho_{de}$ and $H$, we finally obtain
\begin{equation}
w_{de}=\dfrac{1}{3}\left[1-2\beta-\eta\Omega_{de}^\frac{1}{2(2-\beta)}(1-\Omega_{de})^\frac{\beta-1}{2(2-\beta)}e^{\frac{3(1-\beta)}{2(\beta-2)}x}\right] .
\end{equation}
Once again, for $\beta=1$ the above expression reduces to the one of standard HDE model \cite{Li04}.

Moreover, an interesting quantity to consider is the deceleration parameter:
\begin{equation}
q\equiv-1-\dfrac{\dot H}{H^2}=-1+\dfrac{3}{2}(1+w_{de}\Omega_{de})\ ,
\label{eq:deceleration}
\end{equation}
which measures the rate of cosmic expansion, namely decelerating universe for $q>0$ and accelerating universe for $-1\leq q<0$. 
In particular, the transition between the two epochs occurs at $q=0$ .

\section{Cosmological perturbations}
\label{sec:perturbations}

We investigate the theory of linear perturbations in the HDE framework by considering scalar fluctuations of the metric in the Newtonian gauge \cite{Mukhanov92}:
\begin{equation}
ds^2=-(1+2\phi)dt^2+a^2(t)(1-2\phi)\delta_{ij}dx^idx^j\ ,
\end{equation}
where $\phi$ is the Bardeen potential.
Introducing the density contrasts $\delta_i \equiv \delta\rho_i/\rho_i$ and the divergences of the fluid velocities $\theta_i\equiv \vec{\nabla}\cdot \vec{v}_i$, we can write the system of evolution equations for matter and dark energy perturbations in the Fourier space \cite{Abramo09,Mehrabi15}:
\begin{align}
&\ddot{\phi}+4H\dot{\phi}+\left(2\dfrac{\ddot{a}}{a}+H^2\right)\phi=\dfrac{3}{2}H^2c_\text{eff}^2\Omega_{de}\delta_{de}\ , \\
&\dot{\delta}_m+\dfrac{\theta_m}{a}-3\dot{\phi}=0\ , \label{eq:deltamdot}\\
&\dot{\delta}_{de}+(1+w_{de})\left(\dfrac{\theta_{de}}{a}-3\dot{\phi}\right)+3H(c_\text{eff}^2-w_{de})\delta_{de}=0\  \label{eq:deltadedot},\\
&\dot{\theta}_m+H\theta_m-\dfrac{k^2\phi}{a}=0\ , \label{eq:thetamdot} \\
&\dot{\theta}_{de}+H(1-3c_\text{ad}^2)\theta_{de}-\dfrac{k^2c_\text{eff}^2}{(1+w_{de})a}\delta_{de}-\dfrac{k^2\phi}{a}=0\ \label{eq:thetadedot}.
\end{align}
Here, $c^2_\text{eff}\equiv\delta p_{de}/\delta\rho_{de}$ is the effective sound speed, while $c_\text{ad}^2\equiv\dot{p_{de}}/\dot{\rho_{de}}$ is the dark energy adiabatic sound speed:
\begin{equation}
c_\text{ad}^2=w_{de}-\dfrac{aw_{de}'}{3(1+w_{de})}\	 ,
\label{eq:adiabatic}
\end{equation}
where the `prime' denotes derivative with respect to the scale factor.
For $c_\text{eff}^2\simeq 1$, dark energy perturbations are suppressed by pressure and cannot grow on sub-horizion scales, while 
for $c_\text{eff}^2\ll1 $ dark energy and dark matter cluster in a similar way and this affects the growth of structure formation \cite{Erickson02,Linder03,Bean04,Ballesteros08,dePutter10}.
In what follows, we analyze the case of a matter-dominated universe, where $\phi$ is a constant.

At sub-horizon scales $(k^2\gg a^2H^2)$, the Poisson equation reads
\begin{equation}
k^2\phi=-4\pi Ga^2(\delta \rho_m+\rho_{de}+3\delta p_{de})\ ,
\end{equation}
which can be rewritten as
\begin{equation}
-\dfrac{k^2\phi}{a^2}=\dfrac{3}{2}H^2\left[\Omega_m\delta_m+(1+3c^2_\text{eff})\Omega_{de}\delta_{de}\right] . \label{eq:k^2}
\end{equation}
Therefore, we use \Cref{eq:thetamdot,eq:thetadedot} to eliminate $\theta$ from \Cref{eq:deltamdot,eq:deltadedot} by means of \Cref{eq:k^2} to eliminate the $k^2\phi$ terms.
Then, from \Cref{eq:adiabatic,eq:deceleration} with the help of the relation $\frac{d}{dt}=aH\frac{d}{da}$, we obtain the evolution equations for the dark matter and dark energy perturbations:
\begin{align}
& \delta''_m+A_m\delta_m'+B_m\delta_m=S_m\ , \label{eq:matter} \\
& \delta_{de}''+A_{de}\delta_{de}'+B_{de}\delta_{de}=S_{de} \  \label{eq:dark energy},
\end{align}
where 
\begin{subequations}
\begin{align}
&A_m=\dfrac{3}{2a}(1-w_{de}\Omega_{de})\  , \\
&B_m=0\ ,\\
&S_m=\dfrac{3}{2a^2}\left[\Omega_m\delta_m+(1+3c_\text{eff}^2)\Omega_{de}\delta_{de}\right] , \\
&A_{de}=\dfrac{1}{a}\left[\dfrac{3}{2}(1-w_{de}\Omega_{de})-\dfrac{aw_{de}'}{1+w_{de}}-3w_{de}\right] , \\
&B_{de}=\dfrac{1}{a^2}\left[3\left(\dfrac{1}{2}-\dfrac{3}{2}w_{de}\Omega_{de}-\dfrac{aw_{de}'}{1+w_{de}}-3c_\text{eff}^2\right)\right. \nonumber \\	  
&\hspace{1.8cm}\times(c_\text{eff}^2-w_{de})\left.-3aw_{de}'+\dfrac{k}{a^2H^2}c_\text{eff}^2\right] , \\
&S_{de}=\dfrac{3}{2a^2}(1+w_{de})\left[\Omega_m\delta_m+(1+3c_\text{eff}^2)\Omega_{de}\delta_{de}\right] .
\end{align} 
\end{subequations}
From the perturbed Einstein equations, it easy to verify that the following initial conditions hold:
\begin{align}
&\delta_{m}^\text{(ini)}=-2\phi_\text{ini}\left(1+\dfrac{k^2}{3a_\text{ini}^2H_\text{ini}^2}\right) ,\\
&\delta_{m}^{\text{(ini)}'}=-\dfrac{2}{3}\dfrac{k^2\phi_\text{ini}}{a_\text{ini}^2H_\text{ini}^2} \ , 
\end{align} 
while, from the adiabaticity condition, one obtains \cite{Kodama84}
\begin{align}
&\delta_{de}^\text{(ini)}=(1+w_{de})\delta_{m}^\text{(ini)}\ , \\
&\delta_{de}^{\text{(ini)}'}=(1+w_{de})\delta_{m}^{\text{(ini)}'}+w_{de}'\delta_{m}^\text{(ini)}\ .
\end{align}

In this work, we restrict our analysis to the scenario of homogeneous THDE $(\delta_{de}=0,\ c^2_\text{eff}= 1)$ in which the clustering is due only to the corresponding matter component.
Hence, the equation governing the linear matter fluctuations on sub-horizon scales can be written as
\begin{equation}
\delta_m''+\left(\dfrac{3}{a}+\dfrac{E'}{E}\right)\delta_m'-\dfrac{3(1-\Omega_{de})}{2a^2}\delta_m=0 \ ,
\label{eq:growth}
\end{equation}
where $E(a)\equiv H(a)/H_0$ is obtained from \Cref{eq:1/Ha} and $\Omega_{de}(a)$ is given after  solving \Cref{eq:evolution THDE}.

\section{Datasets}
\label{sec:datasets}

To study the observational viability of THDE, we implemented a Bayesian analysis on low-redshift cosmic data, such as type Ia Supernovae (SN) and observational Hubble data (OHD), combined with the growth rate factor (GRF) data of matter fluctuations. We present below the main features of these datasets and describe how they can be used to get bounds over the free parameters of the model presented in the previous section.

\subsection{Supernovae Ia}

The unique leverage offered by SN Ia to investigate the late-time cosmic expansion is testified by the large number of SN surveys over the last two decades probing from very low redshifts ($0.01<z<0.1$) up to $z>1$ \cite{Hicken09,Conley11,Suzuki12}. 
The most recent Pantheon compilation has been presented in \cite{Scolnic17} and consists of a full sample of spectroscopically confirmed PS1 SN Ia previously cross-correlated in \cite{Scolnic15}. 
In the Pantheon dataset, each SN is standardized by means of the SALT2 light-curve fitter \cite{Guy07}, which models the distance modulus as \cite{Betoule14}
\begin{equation}
\mu=m_B-M+\alpha x_1-\beta c + \Delta_M +\Delta_B\ ,
\end{equation}
where $m_B$ is the apparent magnitude of the SN, and $x_1$ and $c$ are the stretch and colour factors of the light curve, respectively;
the $\Delta_M$ term accounts for the host-mass galaxy correction, while $\Delta_B$ is the distance bias correction. In this parametrization, $\alpha$, $\beta$, $M$ and $\Delta_M$ are all nuisance parameters to be determined by fitting the data. 
On the other hand, the cosmological distance modulus is defined as
\begin{equation}
\mu(z)=5\log_{10}\left[\dfrac{d_L(z)}{1\text{ Mpc}}\right]+25\ ,
\end{equation}
where $d_L(z)$ is the luminosity distance which depends on the cosmological parameters of the assumed model. In a flat FLRW universe, this reads
\begin{equation}
d_L(z)=(1+z)\int_0^z\dfrac{dz'}{H(z')}\ .
\end{equation}
The full Pantheon dataset has been used in \cite{Riess18} to construct 6 model-independent\footnote{The $E^{-1}(z)$ measurements rely on the only assumption of a flat universe, which is consistent with our working hypothesis.} and robust $E^{-1}(z)$ measurements, which we utilized in our dark energy analysis.
In this approach, all the SN nuisance parameters are properly marginalized over in the fit. We refer the reader to Table 6 of \cite{Riess18} for the $E^{-1}(z)$ measurements with the corresponding correlation matrix, and for the details of the method.
One can thus write the likelihood probability function of the SN data as
\begin{equation}
\mathcal{L}_\text{SN}\propto \exp\left[-\dfrac{1}{2}\mathbf{A}^\text{T} \mathbf{C}_\text{SN}^{-1} \mathbf{A}\right] ,
\end{equation}
where the vector  $\mathbf{A}$ is given by the difference between the measurements $E^{-1}_i$ and the corresponding values provided by the theoretical model:
\begin{equation}
\mathbf{A}=E^{-1}_{i,obs}-E^{-1}_{th}(z_i) \ , 
\end{equation}
and $\mathbf{C}_\text{SN}^{-1}$ is the inverse covariance matrix built from the correlations between the data points.

\subsection{Observational Hubble data}

The differential age method \cite{Jimenez02} represents a reliable model-independent approach to measure the evolution of the dark energy equation of state at redshifts $z\lesssim 2$, where the universe enters the dark energy-dominated phase. 
This method is based on the spectroscopic dating of galaxy ages, which act as a `clock' measuring the redshift variation of the universe's age.
From the age difference between pairs of nearby passively-evolving galaxies, one can infer the quantity $dz/dt$ and, hence, measure the Hubble parameter according to
\begin{equation}
H(z)=-\dfrac{1}{(1+z)}\dfrac{dz}{dt}\ .
\end{equation}
In our analysis, we used the 31 OHD measurements collected in \cite{solid}. In this case, the data points are uncorrelated, so that the likelihood function reads
\begin{equation}
\mathcal{L}_\text{OHD}\propto\exp\left[-\dfrac{1}{2}\displaystyle{\sum_{i=1}^{31}}\left(\dfrac{H_{obs,i}-H_{th}(z_i)}{\sigma_{H,i}}\right)^2\right]  ,
\end{equation}
where $\sigma_{H,i}$ are the $1\sigma$ uncertainties on the measurements.

\subsection{Growth rate factor}

The large number of dark energy models proposed over the last twenty years has made necessary the study of density inhomogeneities, besides the background evolutionary dynamics, to discriminate among different cosmologies. 
In particular, the growth rate of matter density perturbations is measured through the  quantity $f(a)\equiv a \delta_m'(a)$. 
Redshift-space distortion observations \cite{Macaulay13,Alam17} in the interval $0<z<2$ provide measurements of the factor
\begin{equation}
f\sigma_8(a)\equiv f(a)\sigma_8(a)\ ,
\end{equation} 
where $\sigma_8(a)=\sigma_8\delta_m(a)/\delta(1)$ estimates the linear-density field fluctuations within a $8h^{-1}\text{Mpc}$ radius, with $\sigma_8$ being its current value.
In our work, we considered the Gold-2017 dataset of 18 uncorrelated $f\sigma_8$  measurements presented in \cite{Nesseris17}. These measurements can be used to constrain a specific model only after applying a rescaling procedure with respect to the assumed fiducial cosmology. To this end, we define the ratio
\begin{equation}
r(z)=\dfrac{H(z)d_A(z)}{H_{fid}(z)d_{A,fid}(z)}\ ,
\label{correction}
\end{equation}
where, in this case, the subscript `\emph{fid}' refers to the fiducial $\Lambda$CDM model characterized by the following Hubble expansion rate:
\begin{equation}
H_{fid}(z)=H_0\sqrt{\Omega_{m0}(1+z)^3+(1-\Omega_{m 0}})\ .
\end{equation}
One can thus ``correct'' the measurements by means of the vector
\begin{equation}
\mathbf{Y}=r(z_i)f\sigma_8^{obs}(z_i)-{f\sigma_8^{th}(z_i)}\ ,
\end{equation}
so that the observed values are rescaled with respect to the fiducial model.
Therefore, the likelihood function is given as
\begin{equation}
\mathcal{L}_\text{GRF}\propto \exp\left[-\dfrac{1}{2}\mathbf{Y}^\text{T} \mathbf{C}_\text{GRF}^{-1} \mathbf{Y}\right] ,
\end{equation}
where $\mathbf{C}_\text{GRF}$  is the covariance matrix of the data points (see \cite{Nesseris17} for the numerical values).

\section{Observational constraints}
\label{sec:constraints}

We obtained observational constraints over the cosmological parameters of THDE through a Bayesian analysis on the combined likelihood
\begin{equation}
\mathcal{L}_\text{tot}=\mathcal{L}_\text{SN}\times\mathcal{L}_\text{OHD}\times \mathcal{L}_\text{GRF}\ .
\end{equation}
We notice that the majority of the measurements employed in the present study, namely the SN and GRF data, are insensitive to $H_0$.
On the other hand, the inability of the OHD measurements alone to provide a tight constraint on the Hubble constant would cause difficulties in getting proper bounds over the other cosmological parameters.
For these reasons, in our analysis we fixed the Hubble constant to the most recent best-fit value obtained by the Planck collaboration \cite{Planck18}, $H_0=(67.4\pm0.5)$ km s$^{-1}$Mpc$^{-1}$.

\begin{figure*}
\begin{center}
\includegraphics[width=0.65\textwidth]{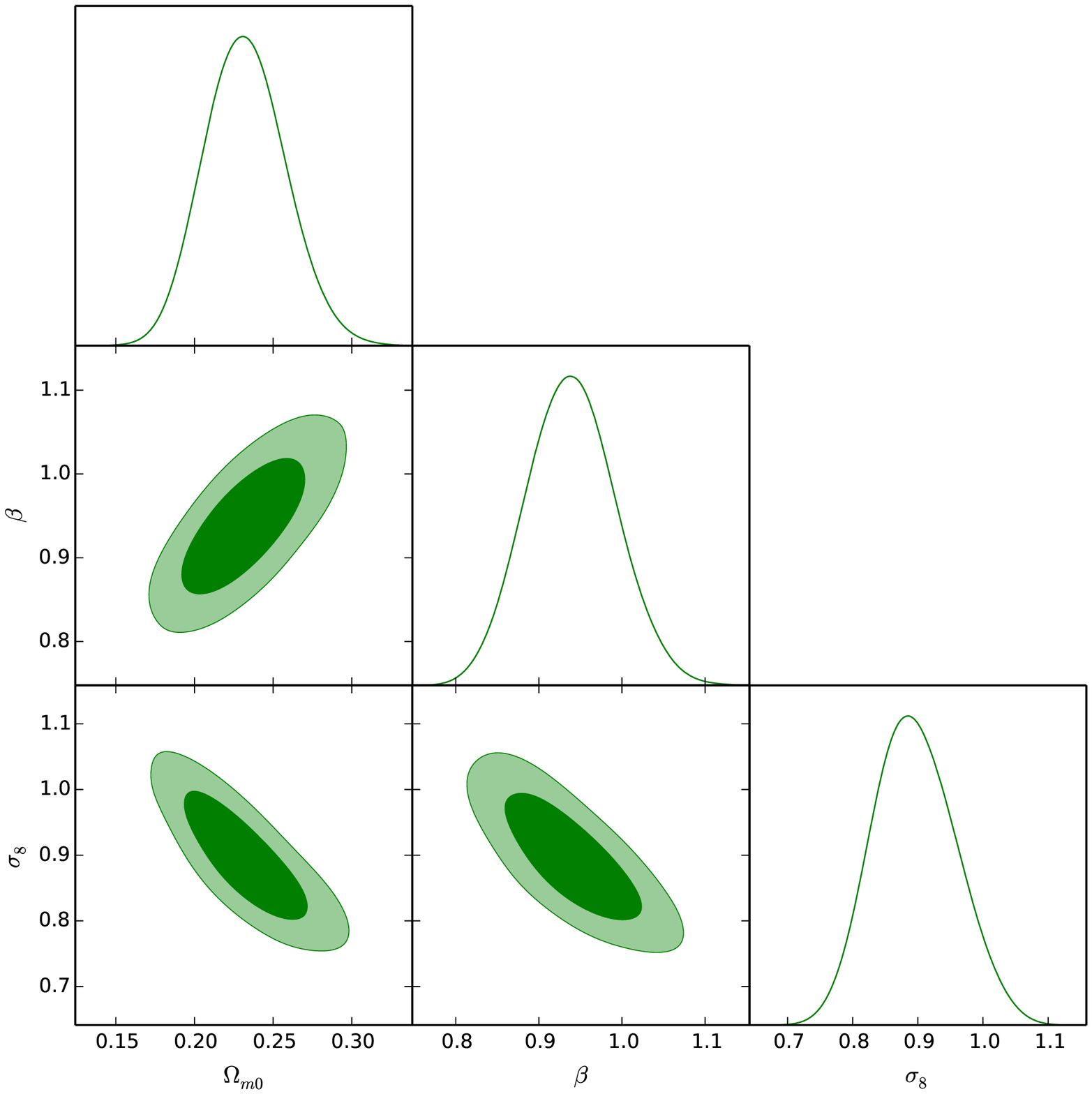}
\includegraphics[width=0.65\textwidth]{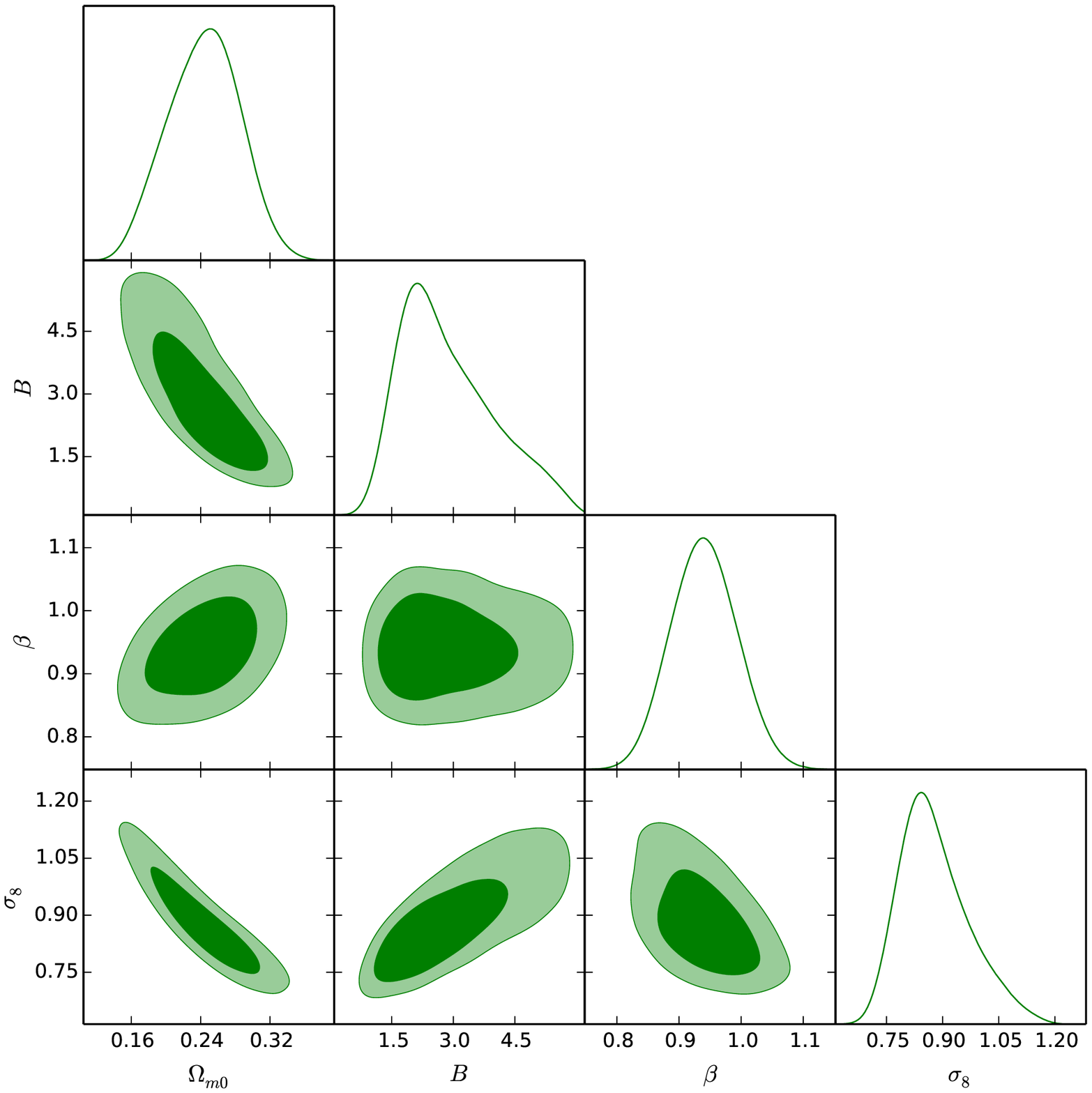}
\caption{Marginalized contours at 68\% and 95\% confidence levels with posterior distributions resulting from the MCMC analysis on THDE models with $B=3$ (top panel) and free $B$ (bottom panel).}
\label{fig:contours}
\end{center}
\end{figure*}

\begin{table*}
\begin{center}
\setlength{\tabcolsep}{1em}
\renewcommand{\arraystretch}{2}
\begin{tabular}{c c c c c c c c c}
\hline
\hline
Model & $\Omega_{m0}$ & $B$  & $\beta$ & $\sigma_8$ & $\Delta$AIC & $\Delta$DIC\\
\hline
THDE 1 & $ 0.232_{-0.027(0.048)}^{+0.024(0.052)} $ & 3 & $0.939_{-0.054(0.101)}^{+0.053(0.107)}$ & $0.895^{+0.060(0.129)}_{-0.069(0.117)}$ & $-1.32$ & $-0.10$   \\
THDE 2 & $0.244_{-0.041(0.081)}^{+0.044(0.079)}$ & $2.864 _{-1.454(1.821)}^{+0.741(2.405)}$ & $0.941_{-0.054(0.101)}^{+0.053(0.104)}$ & $0.879^{+0.066(0.195)}_{-0.111(0.162)} $   & $0.20$ & $1.51$ \\
$\Lambda$CDM & $0.300_{-0.019(0.037)}^{+0.019(0.039)}$ & - & - &  $0.768_{-0.031(0.062)}^{+0.031(0.065)}$ & 0 & 0 \\
\hline
\hline
\end{tabular}
\caption{Mean values and 68\% (95\%) confidence level uncertainties of the cosmological parameters resulting from the MCMC analysis of Tsallis holographic dark energy models with fixed $B=3$ (THDE 1) and free $B$ (THDE 2). The $\Delta$AIC  and $\Delta$DIC values are calculated with respect to the $\Lambda$CDM model, whose results are shown for comparison. Negative values of  $\Delta$AIC and $\Delta$DIC indicate statistical preference for Tsallis holographic dark energy over the $\Lambda$CDM scenario.}
 \label{tab:results}
\end{center}
\end{table*}

\subsection{Monte Carlo analysis}

We obtained the function $\Omega_{de}(a)$ by numerically integrating \Cref{eq:evolution THDE} from the matter-dominated era,  $a_\text{ini}=10^{-2}$, to the present epoch, $a_0=1$, with initial condition $\Omega_{de}(a_0)=1-\Omega_{m0}$.
We thus used this result in \Cref{eq:1/Ha} to find $H(a)$. 
Then, we integrated \Cref{eq:growth} by setting the initial conditions as $\delta_m(a_\text{ini})=10^{-2}$ and $\delta_m'(a_\text{ini})=1$.

Therefore, we applied Markov Chain Monte Carlo (MCMC) method through the Metropolis-Hastings algorithm \cite{Metropolis53}.
In our study, we analyzed two different cases.
In the first scenario, we fixed $B=3$ to obtain exact correspondence with standard HDE model and look for possible deviations from $\beta=1$. Hence, we performed the sampling over the parameter space 
\begin{equation}
\mathcal{P}_1=\{\Omega_{m0}, \beta,  \sigma_8\} .
\end{equation}
In the second scenario, we left $B$ as a free parameter in the numerical procedure to enhance the capability of THDE to explain cosmological observations. We thus sampled  over the following parameter space:
\begin{equation}
\mathcal{P}_2=\{\Omega_{m0}, B, \beta, \sigma_8\} .
\end{equation}
The sampling has been done assuming uniform priors for the cosmological parameters:
\begin{equation}
\left\{
\begin{aligned}
&\Omega_{m0}\in(0,1)\ ,\\
&B\in(0,6)\ , \\
&\beta\in(0.5,1.5) \ , \\
&\sigma_{8}\in(0.5,1.5)\ .
\end{aligned}
\right.
\end{equation}
We implemented our numerical code by means of the software \texttt{Mathematica}, while we used the \texttt{getdist}\footnote{\url{https://getdist.readthedocs.io}} package to analyze the chains and produce the contour plots.

We report in \Cref{tab:results} the $1\sigma$ and $2\sigma$ results for the cosmological parameters of the THDE models under consideration.
Moreover, we display in \Cref{fig:contours} the corresponding two-dimensional marginalized confidence level contours and the one-dimensional posterior distributions.  
The results for the first model show that there is no significant deviation from the standard HDE model with $\beta=1$. In fact, the two scenarios are only slightly more than $1\sigma$ away from each other, in agreement with the previous findings of \cite{Saridakis18}.
This behaviour is confirmed even when $B$ is allowed to vary. In this case, the data are not able to provide tight constrains on $B$, whose estimate is consistent with the value of the standard HDE model $(B=3)$.

\begin{figure}
\begin{center}
\includegraphics[width=3.3in]{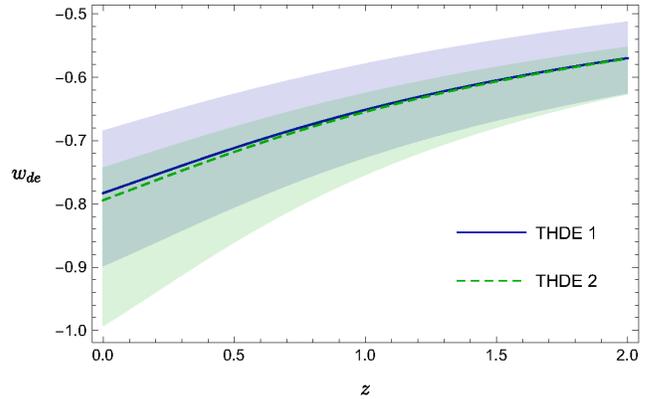}
\caption{Equation of state parameter for Tsallis holographic dark energy models with fixed $B=3$ (THDE 1) and free $B$ (THDE 2) resulting from our MCMC analysis. The shaded regions around the mean curves take into account the $1\sigma$ errors on $\beta$ in the case of THDE 1 and on $B$ in the case of THDE 2, while the other cosmological parameters are fixed to the mean values.}
\label{fig:wDE}
\end{center}
\end{figure}

In \Cref{fig:wDE} we show the evolution of THDE equation of state parameter assuming the numerical values obtained from the MCMC analysis. Our results indicate a quintessence-like behaviour and no phantom-divide crossing is found within the $1\sigma$ confidence level.
It is interesting to compare our $w$ values with the results found in \cite{Scolnic17} combining the CMB and the SN data, which are consistent with the cosmological constant case. We note that our outcomes are compatible with those only in the case of THDE model with free $B$ at the lower $1\sigma$ bound. As one can see from \Cref{tab:results}, the evidence for $w>-1$ in our models is compensated by a shift of the matter densities towards lower values with respect to the standard scenario, characterized by $\Omega_{m0}\approx 0.3$.

\begin{figure}
\begin{center}
\includegraphics[width=3.2in]{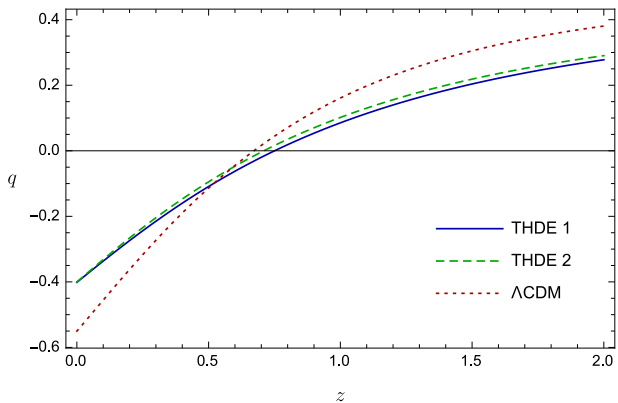}
\caption{Evolution of the deceleration parameter for Tsallis holographic dark energy models with fixed $B=3$ (THDE 1) and free $B$ (THDE 2). We assumed the mean values of the cosmological parameters resulting from our MCMC analysis. The $\Lambda$CDM curve is shown for comparison. The points where the curves intersect the black line indicate the transition from decelerated to accelerated universe.}
\label{fig:deceleration}
\end{center}
\end{figure}

\begin{figure}
\begin{center}
\includegraphics[width=3.2in]{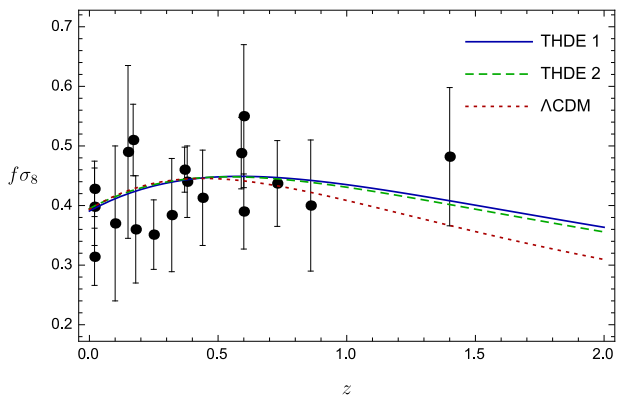}
\caption{Growth of matter overdensities for Tsallis holographic dark energy models with fixed $B=3$ (THDE 1) and free $B$ (THDE 2). We assumed the mean values of the cosmological parameters resulting from our MCMC analysis. The $\Lambda$CDM curve is shown for comparison.}
\label{fig:growth}
\end{center}
\end{figure}

We also show in \Cref{fig:deceleration} the behaviour of the deceleration parameter with the redshift. 
Furthermore, in \Cref{fig:growth} we display the growth rate of matter fluctuations for the THDE models compared to the $\Lambda$CDM scenario.

Finally, to estimate the impact of our \textit{a priori} assumption on $H_0$, we repeated the statistical analysis by fixing the Hubble constant to values that are $1\sigma$ away from the central value obtained in \cite{Planck18}. In the case of the THDE model with fixed $B=3$, we found values of the cosmological parameters that differ only by $1-2\%$ from the results shown in \Cref{tab:results}. In the case of the THDE model with varying $B$, it turned out that the differences with respect to the previous value of $B$ amount to $\sim5\%$ with similar relative uncertainties, while the other cosmological parameters are different, once again, only by $1-2\%$. These outcomes demonstrate the robustness of our results and the accuracy of our procedure.

\subsection{Bayesian model selection}

We measured the statistical evidence of the THDE models by means of information criteria estimators. In particular, we used the Akaike information criterion (AIC) 	\cite{Akaike74}:
\begin{equation}
\text{AIC}\equiv-2\ln\mathcal{L}_\text{max}+2p\ ,
\end{equation}
where $\mathcal{L}_\text{max}$ is the maximum value of the likelihood and $p$ is the number of parameters of the theoretical model.
We also considered the DIC criterion, which accounts for the number of parameters that can be effectively constrained by a specific dataset \cite{Kunz06}:
\begin{equation}
\text{DIC}\equiv\langle-2\ln\mathcal{L}\rangle+p_\text{eff}\ ,
\end{equation}
where $p_\text{eff}=\langle-2\ln\mathcal{L}\rangle+2\ln\langle\mathcal{L}\rangle$, with $\langle\cdot\rangle$ denoting average over the posterior distribution.
The best model is the one characterized by minimum AIC and DIC values. 

In our case, we computed AIC and DIC differences with respect to the $\Lambda$CDM model, chosen as the reference model.
The results shown in \Cref{tab:results} indicate that the THDE model with fixed $B=3$ performs better than $\Lambda$CDM, albeit with poor significance level.
On the other hand, allowing $B$ to vary in the fitting procedure does not increase the Bayesian evidence for the corresponding THDE model, which appears penalized by the presence of the extra parameter.

\section{Final outlook and perspectives}
\label{sec:conclusion}

In the present work, we assumed the validity the holographic principle to address the accelerated cosmic expansion. 
In particular, we discussed the features of a dark energy model built upon the Tsallis entropy, which represents a nonadditive generalization of the Boltzmann-Gibbs entropy that should be used in the statistical treatments of nonextensive systems, such as gravitational ones. 
We considered the future event horizon as the characteristic length of the universe. 
This choice allowed us to avoid cosmological inconsistencies and include standard holographic dark energy and standard thermodynamics as sub-classes.
 
We described the dynamics of Tsallis holographic dark energy at the level of background cosmology. 
Furthermore, we studied linear perturbations on a flat FLRW spacetime at sub-horizon scales.  
We thus focused on the case of homogeneous dark energy in a matter-dominated universe and derived the growth rate of matter density fluctuations.

We analyzed two specific THDE scenarios: a first model with fixed $B=3$, which recovers standard HDE in the limit $\beta=1$, and a second model with $B$ left as a free parameter. 
We thus tested the observational viability of the aforementioned theoretical models by a comparison with the most recent cosmological data. Assuming uniform priors for the free parameters, we performed MCMC numerical technique implemented through the Metropolis-Hastings algorithm on the combined likelihood of SN Ia data, Hubble parameter measurements and growth rate factor data. 
In the first scenario, we found that the deviation from standard HDE is within the $2\sigma$ level. 
In the second scenario, our results show the inability of the data to provide tight constraints on the parameter $B$ and, consequently, no deviation from the value of the standard HDE model was found.
Then, assuming the numerical results of the MCMC analysis for the cosmological parameters, we computed the redshift evolution of the dark energy equation of state and the deceleration parameter. We showed that the equation of state of THDE models behaves as quintessence. We also compared the growth rate of matter overdensities with the predictions of the concordance $\Lambda$CDM paradigm.

Moreover, we used the AIC and DIC information criteria to measure the Bayesian evidence for the models under consideration. We found that the first THDE scenario performs slightly better than the $\Lambda$CDM model, while the second THDE scenario is statistically penalized by the additional free parameter. 

Possible extensions of this work include the use of high-redshift data, such as Cosmic Microwave Background observations, to complement the constraints at late times. It would be also interesting to consider the effects of clustering dark energy with $c_\text{eff}^2\simeq 0$ on the evolution of matter perturbations. We leave these issues for future investigations.

\begin{acknowledgements}

The author is grateful to Orlando Luongo and Constantino Tsallis for useful comments and suggestions.

\end{acknowledgements}

\end{document}